\documentclass[10pt]{article}
\usepackage{a4}
\usepackage{epsfig}
\newcommand{\Ut}{U_{\rm tot}}
\newcommand{\nn}{\nonumber}
\newcommand{\pdif}[3]{\left( \frac{\partial #1 }{\partial #2 }\right)_{#3 }}  
\newcommand{\pd}[2]{\frac{\partial #1 }{\partial #2 }}  
  
\newcommand{\midd}[1]{\langle #1 \rangle}
\newcommand{\be}{\begin{equation}}
\newcommand{\ee}{\end{equation}}
\newcommand{\BE}{\begin{eqnarray}}
\newcommand{\EE}{\end{eqnarray}}
\newcommand{\x}{\vec{p}_1,\ldots, \vec{p}_N;\vec{q}_1,\ldots, \vec{q}_N} 
\newcommand{\dx}{\,d\vec{p}_1\cdots d\vec{p}_N\,d\vec{q}_1\cdots d\vec{q}_N} 
\newcommand{\CR}{Cram\'er-Rao inequality }
\newcommand{\forget}[1]{}

\newcommand{\ket}[1]{| #1 \rangle}
\newcommand{\bra}[1]{\langle #1 |}

\newcommand{\inpr}[2]{\langle #1 | #2 \rangle}
\newcommand{\ges}{%
\raisebox{-.6ex} 
{$\mbox{}\,\stackrel{\displaystyle>}{\displaystyle\sim}\,$ }}

\title{Thermodynamic uncertainty relations}
\author{Jos Uffink and Janneke van Lith-van Dis\thanks{Institute for
History and Foundations of Mathematics and Science,
University of Utrecht, P.O.Box 80.000, 3508 TA Utrecht, The
Netherlands ({\tt uffink@phys.uu.nl, j.h.vanlith@phys.uu.nl})
}}
\begin{document}
\maketitle
\begin{abstract}
Bohr and Heisenberg suggested that the thermodynamical quantities of
temperature and energy are complementary in the same way as position and
momentum in quantum mechanics.  Roughly speaking, their idea was that a
definite temperature can be attributed to a system only if it is submerged
in a heat bath, in which case energy fluctuations are unavoidable. On
the other hand, a definite energy can only be assigned to systems in
thermal isolation, thus excluding the simultaneous determination of its
temperature. 

Rosenfeld extended this analogy with quantum mechanics and obtained a
quantitative uncertainty relation in the form $\Delta U \Delta (1/T)  \geq
k$ where $k$ is Boltzmann's constant.  The two `extreme' cases of this
relation would then characterize this complementarity between isolation
($U$ definite) and contact with a heat bath ($T$ definite).
  Other formulations of the thermodynamical uncertainty relations were
proposed by Mandelbrot (1956, 1989), Lindhard (1986) and Lavenda (1987,
1991). This work, however, has not led to a consensus in the literature. 

It is shown here that the uncertainty relation for temperature and energy
in the version of Mandelbrot is indeed exactly analogous to modern
formulations of the quantum mechanical uncertainty relations.  However,
his relation holds only for the canonical distribution, describing a
system in contact with a heat bath. There is,
therefore, no complementarity between this situation   and 
a thermally isolated system.
  \end{abstract}

\section{Introduction}

The uncertainty relations and the principle of complementarity are usually
seen as hallmarks of quantum mechanics. However, in some writings of Bohr
and Heisenberg$^{\ref{Bohr}}$ 
 one can
find the idea that there also is a
complementary relationship in classical physics, in particular between the
concepts of energy and temperature.
 Roughly speaking, their argument is that the only way to attribute a
definite temperature to a physical system is by bringing it into thermal 
contact and equilibrium with another very large system acting as a heat bath.  
In this case, however, the system will freely exchange energy with the heat
bath, and one is cut off from the possibility of controlling its energy. 
On the other hand, in order to make sure a system has a definite energy,
one should isolate it from its environment. But then there is no way to
determine its temperature.

 This idea is in remarkable analogy to Bohr's famous analysis of
complementarity in quantum mechanics, which is likewise based on a similar
mutual exclusion of experimental arrangements serving to determine the
position and momentum of a system. 
 And just as this complementary relationship in quantum mechanics finds
its `symbolic expression', as Bohr puts it, in the uncertainty relation
$\Delta p \Delta q \ges  \hbar$, one might expect to 
obtain an analogous uncertainty relation for energy and 
temperature,  or, perhaps, some functions of these quantities.
Dimensional analysis already leads to the conjecture that this relation
would take the form  \be
\Delta U \Delta \frac{1}{T} \ges k
   \label{ur}
\ee
where  
 $k$ denotes the Boltzmann constant. 

However these ideas have not received much recognizition in the
literature. 
An obvious objection is  that the mathematical structure of quantum theory is 
radically different from that of clasical physical theories.
  There are no non-commuting observables in thermodynamics. 
Therefore, a derivation of the uncertainty relation (\ref{ur}) analogous
to that of
the usual Heisenberg relations is impossible. 
  For example, the   biography  of Bohr 
by Pais$^{\ref{Pais}}$ 
 dismisses his proposal of a complementary principle in thermodynamics on
the grounds that there does not exist a general uncertainty relation like
(1) to back it up. 
 
Nevertheless, several authors$^{\rm 3-12} $
have in fact  produced derivations of a
relation of this kind, and defended  the idea  that  it  reveals a 
complementarity between thermal isolation and embedding in a heat bath. 
An even  more far-reaching claim is put forward by Rosenfeld$^{\rm 6}$,
Mandelbrot$^{\rm 3-5}$  and Lavenda$^{7-10}$.
 These authors argue that the uncertainty relation (\ref{ur}) actually blocks
any reduction of thermodynamics to a microscopic theory picturing an underlying
molecular reality, in the same way as Heisenberg's relation would forbid a
`hidden variables' reconstruction of quantum mechanics.
  
However, these claims, and indeed the very validity of (\ref{ur}),
have been disputed by other physicists. 
An example of this controversy is 
provided by the polemic exchange between Feshbach, Kittel and Mandelbrot 
in the years 1987-9 in \emph{Physics Today}.$^{\rm 13-15}$
One of the immediately arising questions is what the exact meaning of the
$\Delta$'s in (\ref{ur})  could be. A first thought may be that these 
uncertainties are to be understood as standard deviations of a random
quantity, according to one of the probability distributions (or ensembles)
of statistical
mechanics. But in the commonly used ensembles to describe a system in contact
with a heat bath, the canonical ones, temperature is just a fixed parameter 
and doesn't fluctuate at all. On the other hand, in the microcanonical 
ensemble the energy is fixed! 
So this straightforward interpretation cannot be correct. 

Indeed, all versions of uncertainty relation (\ref{ur}) which have been
proposed in the literature employ different theoretical frameworks and
give different interpretations of the uncertainty $\Delta \beta$, in most
cases by using concepts from theories of statistical inference.  Our
purpose is to review and criticize the existing derivations and the
associated claims (Sections 2 and 3).  We shall argue that even though
there exist valid versions of this relation, it does not express
complementarity between energetic isolation and thermal contact.

 We also propose (Section 4) an elaboration of the approach of Mandelbrot,
which is based on the
Cram\'er-Rao inequality from the theory of statistical estimation. 
We generalize this approach by means of the concept of distance between 
probability distributions. This statistical distance is a measure of
distinguishability: the smaller the distance between two probability
distributions, the worse they can be distinguished by any method of
mathematical statistics. 
 It will be shown  that this statistical distance leads to an improvement
of the Cram\'er-Rao inequality.  The resulting uncertainty relation has
the virtue of being completely analogous to a quantum mechanical formulation.
 In section 5 we will use this approach to investigate intermediate cases
between the canonical and microcanonical ensemble.

\section{The approach from fluctuation theory: Rosenfeld and Schl\"ogl} 
The best-known proposal for a quantitative
uncertainty relation between energy and temperature is that of
Rosenfeld$^{\ref{Rosenfeld} }$. 
He obtained the result
 \be 
  \Delta U \Delta T = kT_0^2 
  \label{urRosenfeld}
\ee
 for the standard deviations (root mean square fluctuations)  of energy
and temperature of a small but macroscopic system in thermal contact with
a heat bath kept at the fixed temperature $T_0$. In the case of small 
fluctuations, $\Delta T \ll T_0$,
$\Delta\beta \approx \frac{\Delta T}{kT_0^2}$
and Rosenfeld's result reduces to (\ref{ur}).

Rosenfeld's interpretation of relation (\ref{urRosenfeld})  fits
seamlessly into the Copenhagen tradition. He argues that the meaning of
physical quantities has to be obtained by operational definitions, 
i.e.\ by referring to our experimental abilities to control their values.
 Thus the meaningfulness of quantities depends on the experimental
context, just like in quantum mechanics. Energetic isolation and thermal
contact with a heat reservoir are contexts that exclude each other and
therefore $U$ and $T$  are not simultaneously meaningful.
 The uncertainty relation (\ref{urRosenfeld}) symbolizes this fact. 
More generally, Rosenfeld  speaks of a complementarity 
 between dynamical and thermodynamical modes  of description.
 
This complementarity, according to Rosenfeld, is equally fundamental as
Heisenberg's uncertainty relation.  He argues that reference to the
underlying microscopic constitution of the system will not succeed in
restoring a unified description, just because the dynamical and
thermodynamical concepts are only meaningful in mutually exclusive
experimental conditions. 

For the actual derivation of (\ref{urRosenfeld}) Rosenfeld refers to the
textbook of Landau and Lifshitz$^{\ref{LL}}$.
 This book offers a treatment of thermal fluctuations for a small macroscopic
system in an environment at fixed temperature $T_0$ and pressure $p_0$. It
is based on Einstein's postulate, which inverts and reinterprets
Boltzmann's famous formula
$
  S=k\ln W
$
  into:
\be
 W(X) \propto e^{S_{\rm tot}(X)/k} 
\ee
 in order to assign relative probabilities  to thermodynamical states $X$ in
terms of their entropy $S$. Here, $X$ denotes a state of the total system
(small system and environment) and $S_{\rm tot}$ is the total entropy. 
 An equivalent formulation is
 \be W(X)  = W(X_0)  e^{(S_{\rm tot}(X) - S_{\rm tot}(X_0))/k}
 \label{ep}
\ee 
 where $X_0$ is the equilibrium state. 

The probabilities $W(X)$ are interpreted as the relative frequencies with
which the states $X$ occur during a very long time interval, i.e.\ they
describe fluctuations.
 An essential assumption is now that both the small system and the
environment taken separately are always in thermodynamical equilibrium
states. Thus, even though they need not be in thermal or mechanical
equilibrium with each other, it is assumed that they can always be 
characterized by equilibrium states on which the ordinary
thermodynamical formalism is applicable. 
 The entropy fluctuation $S_{\rm tot}(X) -S_{\rm tot}(X_0)$ of the total
system can then be calculated in terms of the minimal amount of work
needed to restore the equilibrium state $X_0$. This can be expressed in
the state variables of the small system and the fixed equilibrium values
$p_0$ and $T_0$: 
\be  \label{stotx} 
 S_{\rm tot}(X) -S_{\rm tot}(X_0) =- \frac{U(x)  +p_0 V}{T_0} + S(x) 
\ee
   where the variable $x$ denotes the state of the small system.  
Thus, from (\ref{ep}) and (\ref{stotx}) one can determine the
fluctuation probabilities in terms of
the  state variables of the small system. Choosing a complete 
set of independent state variables, say $x =(V,T)$, and expanding (\ref{ep}) 
up to second order around its maximum value, one can approximate this
distribution by a Gaussian probability distribution over $V$ and $T$.
  Landau and Lifshitz determine the standard deviations of several
thermodynamic quantities considered as functions of $V$ and $T$.  
In particular they obtain (pp.~352, 356)
 \be  
(\Delta  T)^2   = \frac{kT_0^2}{C_V} \label{DT}\ee
and 
 \be  \label{DU}
(\Delta  U)^2   = -  \left( T_0 \pdif{P}{T}{V} - p_0 \right)^2 kT_0 
\pdif{V}{P}{T}   +  C_V kT_0^2.
\ee
Here, $C_V =\pdif{U}{T}{V}$  is the specific heat of the system
and the quantities in the right-hand sides of (\ref{DT},\ref{DU}) 
are all to be evaluated at the equilibrium values. 
Finally,
in order to arrive at (\ref{urRosenfeld}),
Rosenfeld 
simply assumed that the volume of the system is
constant (so that $\pd{V}{P} =0$). 

Let us now see if this result bears out Rosenfeld's interpretation.
There are several objections. A first group
of objections is directed in particular against his claim that the result
would be `equally fundamental' as the Heisenberg uncertainty relation.  First,
the relation (\ref{urRosenfeld}) is obtained by ignoring the first term in
the right-hand side of
(\ref{DU}). This is obviously not a satisfactory general procedure, 
unless one can prove that the deleted term is always non-negative. 
 For usual thermodynamical systems, one has
indeed$^{\ref{LL62}}$
\be 
   T_0> 0, \;\;\;\ \pdif{V}{P}{T}<0\label{ei}
\ee
so that a more general inequality
$\Delta U\Delta T \geq kT_0^2$ results from (\ref{DT}) and (\ref{DU}). 
 However, the inequalities (\ref{ei}) are in turn derived by an
appeal to the stability of thermodynamical states. The assumption that 
all equilibrium states occurring in nature are stable is obviously not
a fundamental law.

Secondly, the above derivation relies on a Gaussian approximation to
(\ref{ep}). This, likewise, cannot claim fundamental validity. Moreover,
the question how well the Gaussian distribution approximates the true
distribution (\ref{ep}) depends on the choice of the variables to
parameterize the macroscopic states.  (That is, whether one uses the pairs
$(V, T)$, $(P,V)$ etc., or prefers $T$ above $\beta$, or perhaps
$(T-T_0)^3$ as a temperature scale.)  The choice of state variables is
usually regarded as a matter of convention, and one would not like a
fundamental result to depend on this. 

A third objection is that whatever limitations (\ref{urRosenfeld})  poses
on the simultaneous meaningfulness of the concepts of energy and
temperature, a look at the right-hand side of this relation immediately
suggests that these limitations will become negligible if we take the
temperature of the heat bath sufficiently low. This again is in obvious
contrast to Heisenberg's uncertainty relation which provides an absolute
lower bound for all quantum states.

Remarkably, it is possible to overcome all the above objections and obtain
a more generally valid uncertainty relation for the situation considered
here.  To see this, let us choose $(U,V)$ as the set of independent
state variables and write the probability distribution 
resulting from the Einstein postulate in the form
 \be
p(U,V) = C e^{-\beta_0(U +p_0V) + S(U,V)/k} \label{ep1}\ee
and introduce the quantity
 \be  \beta(U,V)  := \pd{\ln p(U,V)}{U} + \beta_0 \label{b}\ee
It is easy to see that
$\beta(U,V )
  =\pdif{S}{U}{V} $, and is thus identical to the usual thermodynamical
definition of the inverse temperature of the system. Using the
general inequality
\be 
  \Delta A \Delta B \geq |\mbox{Cov}(A,B)|:= \midd{(A- \midd{A})(B -
  \midd{B})} 
\ee
where $\midd{\cdot}$ denotes the expectation value  with respect to
(\ref{ep1}) and noting that $\mbox{Cov}(U,\beta)= -1$ 
 (by partial integration and assuming that $p(U,V) =0$ when $U=0$),
one obtains 
\be \Delta U \Delta \beta \geq 1 \label{urSchloegl}
\ee
independent of stability arguments, Gaussian approximation or the value 
of $T_0$. This relation was first obtained by 
Schl\"ogl$^{\ref{Schloegl}}$.

Let us make a few comments on this improved result.
 Note that there is no assumption of an underlying mechanical phase space
of the system. Rather, one works directly with probability distributions
over the macroscopically observable variables.  Thus, the derivation is
not a part of the Gibbsian theory of statistical mechanics. Rather, the
above treatment, combining orthodox thermodynamics with a probability
postulate is typical for statistical thermodynamics.  In principle, it is
an open question whether this version of statistical thermodynamics is
consistent with the existence of an underlying microscopic phase space on
which all thermodynamical quantities can be defined as functions.
 
Indeed a remarkable aspect of this version of statistical thermodynamics
is that quantities like temperature and entropy depend on the probability
distribution, via the Einstein postulate.  This means one cannot vary the
probability distribution and the temperature independently.  Obviously,
this aspect alone already provides an obstacle to a hidden-variables-style
reconstruction.

The same aspect, however, makes the relation (\ref{urSchloegl}) rather 
different from quantum mechanical uncertainty relations.  
It states only that there are non-vanishing fluctuations in energy  and
inverse temperature for the distribution (\ref{ep1}). For an arbitrary
distribution, the
quantity  $\beta$ defined by  (\ref{b}) does not coincide with
the inverse
temperature. 
This  limited validity does not license  a complementarity interpretation.
For example, for the ideal gas one has $\beta(U,V) =\frac{C_V}{kU}$, 
with $C_V$ a constant, i.e.\  $\beta$ is a bijective function  of $U$.
 Clearly, as 
Lindhard$^{\ref{Lindhard}}$ 
 pointed out, any claim that   
`precise knowledge of $U$ precludes precise knowledge of $\beta$'
would be quite untenable here. 
The positive lower bound for $\Delta U \Delta\beta$ is due, in this
case, to their correlation rather  than  their complementarity.
One is not left, for a given system, with a choice of making $\Delta U$ 
smaller at the expense of $\Delta \beta$ or vice versa. Indeed,
 the relative values of the two uncertainties in (\ref{ep1})  are decided
by the size of the system (which determines the heat capacity $C_V$), 
 not by context of observation. 
  
Thus, important objections against Rosenfeld's interpretation of the
uncertainty relation (\ref{urRosenfeld}) remain, even in the improved
version (\ref{urSchloegl}). Note also that while Rosenfeld speaks of a
complementarity between thermal contact and energetic isolation, the above
treatment nowhere refers to isolated systems. The only experimental
context considered in the derivation is that of thermal contact with a
heat bath. Therefore, the basis for Rosenfeld's assertions is very weak. 

 Perhaps Rosenfeld's interpretation was inspired by a short remark at the
end of Landau and Lifshitz's 
discussion$^{\ref{LL355}}$,
 where they
mention that temperature fluctuations can also be considered ``from
another point of view'', and briefly discuss the canonical and
microcanonical ensembles.  They state that one must assume temperature
fluctuations exist also in an isolated system, so that the result
(\ref{DT}) ``represents the accuracy with which the temperature of an
isolated body can be specified.'' However, none of this follows from the
treatment they actually present. 

Now one could try to provide such an analysis of thermal fluctuations
in an energetically isolated  system along the lines of Landau and Lifshitz. 
 However, this will  obviously lead to  $\Delta U =0$, 
 and in the example of the ideal gas, $\Delta \beta =0$ as well. 
 The only way in which temperature fluctuations can be obtained is
therefore
 either by specializing to systems where $\beta$ is not a function of $U$
 alone (e.g. a photon gas where $\beta(U,V) \propto (V/U)^{1/4}$),
 or by generalizing the approach to allow for local fluctuations within
 different parts of the system. However, we shall not pursue this.

 \section{Statistical inference}
 The approaches to the derivation of thermodynamic uncertainty relations
we discuss next are all related to the field of statistical inference. We
therefore start this section with a short introduction to this subject. 
  The theory of statistical inference has been developed in the
twenties and thirties by mathematical statisticians, working largely
outside the physics community. The physicists working in statistical
physics have for a long time continued thinking about statistics using the
concepts laid down in the older work of Maxwell, Boltzmann and Gibbs. 
Still, it has been recognized since the sixties that statistical physics
can benefit from the ideas and concepts developed in mathematical
statistics.  For our purpose the most significant aspect is that more
sophisticated concepts of uncertainty are available here than the standard
deviation. 
 However, the field of statistical inference is one in which several
approaches exist and there is a longstanding debate about its foundations.
If statistical physics can profit from the developments in statistical
inference, it also cannot remain immune to this debate, as we shall see.
We shall meet four approaches to statistical inference: 
 estimation theory, Bayesianism, fiducial probability and likelihood
inference.  We shall see that the so-called Fisher information plays a
prominent (but different) role in most of them. 

\subsection{Estimation theory} 
 Generally speaking, statistical inference can be described as the problem
of deciding how well a set of outcomes, obtained in independent
measurements, fits to a proposed probability distribution. If the
probability distribution is characterized by one or more parameters, this
problem is equivalent to inferring the value of the parameter(s) from the
observed measurement outcomes $x$.  Perhaps the simplest, and most
well-known approach to the problem is the theory of estimation, developed
by R.A. Fisher. 

 In this approach it is assumed that one out of a family 
$\{p_{\theta}(x) \}$ of distribution functions is the `true' one; the
parameter $\theta$
being unknown.  To make inferences about the parameter, one
constructs estimators, i.e.\
 functions $\hat{\theta}(x_1 , \ldots, x_n)$ of the outcomes of $n$
independent repeated measurements.  The value of this function is intended
to represent the best guess for $\theta$. 
 Several criteria are imposed on these estimators in order to ensure that
their values do in fact constitute `good' estimates of the parameter
$\theta$. \emph{Unbiasedness} for instance, i.e.\
 \be
   \midd{\hat{\theta}}_{\theta}=\int \hat{\theta}(x_1, \ldots, x_n)
    \prod_{i=1}^{n} p_\theta (x_i) dx_i = \theta~~~~~~~~~\mbox{(for all
$\theta$)}
\ee
 demands that the expectation of $\hat{\theta}$, calculated using a given
value of $\theta$, reproduces that value. If also the standard deviation
$\Delta_\theta \hat{\theta}$ of the estimator is as small as possible, the
estimator is called \emph{efficient}.

The so-called Cram\'er-Rao inequality puts a bound to the efficiency
of an arbitrary estimator:
\be
  (\Delta_\theta \hat{\theta})^2 \geq 
  \frac{\left(|\frac{d\midd{\hat{\theta}}_\theta}{d\theta}|\right)^2}
  {n I_F(\theta)}
   \label{CR}
\ee
where
\be
  I_F(\theta)=\int \frac {1}{p_\theta(x) } 
  \left( \frac{d p_\theta(x)}{d\theta}\right)^2  dx
  \label{I_F}
\ee
is a quantity depending only on the family $\{p_\theta(x)\}$,  known as
 the \emph{Fisher information}.

The Cram\'er-Rao inequality (\ref{CR}) 
is valid for any estimator for a given family of probability distributions 
obeying certain regularity conditions$^{\ref{reg}}$.
Specific choices of estimators are the maximum likelihood estimators.
These are functions $\hat{\theta}_{ML}$ which maximize the likelihood
$L_{(x_1, \ldots, x_n)}(\theta) = \prod_{i=1}^n p_\theta(x_i)$.  They
 are asymptotically efficient, i.e.\ they approach the bound of the
Cram\'er-Rao inequality in the limit $n\rightarrow \infty$. 

Another important criterion in estimation theory is that of 
\emph{sufficiency}. Suppose that for a given estimator 
$\hat{\theta}(x_1, \ldots, x_n)$ the
probability distribution function can be written as
\be
   p_\theta(x_1,\ldots,x_n)=
\tilde{p}_\theta(\hat{\theta})
f(x_1,\ldots,x_n)
 \label{suff}       
\ee
where $\tilde{p}_\theta(\hat{\theta})$ is the marginal distribution of 
$\hat{\theta}$  and $f$ is an arbitrary function which does not 
depend on $\theta$.
Thus, given the value of $\hat{\theta}(x_1, \ldots, x_n)$, the values of
the data $x_1,\ldots, x_n$ are  distributed independently of $\theta$.
In that case, $\hat{\theta}(x_1,\ldots, x_n)$ is said to be a 
sufficient estimator, because it contains all the information about the
parameter that can be obtained from the data.

Sufficiency is a natural and appealing demand for estimation problems.
 Unfortunately sufficient estimators do not always exist.
 A theorem by Pitman and Koopman states that sufficient estimators exist
only for the so-called \emph{exponential family}, i.e.\ distributions of
the form 
 \be p_\theta(x) = \exp \left(A(x) + B(x)C(\theta) +
D(\theta)\right) \label{PK}
 \ee 
 where $A, \ldots, D$ are arbitrary functions (apart, of course, from the
normalization constraint).  Fortunately, most of the one-parameter
distributions that we meet in statistical physics do belong to the
exponential family.

Note that the Fisher information  for $\theta$ in $n$ observations remains 
unaffected when we restrict the set of data to a sufficient estimator:
\be   I_F(\theta)=\int \frac {1}{p_\theta(x_1 ,\ldots , x_n) } 
 \left( \frac{d p_\theta(x_1, \ldots, x_n)}{d\theta}\right)^2  
  dx_1 \cdots dx_n = 
 \int \frac {1}{\tilde{p}_\theta(\hat{\theta}) } 
  \left( \frac{d\tilde{p}_\theta(\hat{\theta})}{d\theta}\right)^2  d\hat{\theta}
\ee
where $\tilde{p}_\theta(\hat{\theta})$ is the marginal probability distribution
of $\hat{\theta}$. By contrast,
 $I_F$  decreases  when the data are restricted to a 
non-sufficient estimator.  In this sense too,  a sufficient estimator extracts
the maximum amount of information about $\theta$ from the data. 
 
\subsection{Uncertainty relations from estimation theory: Mandelbrot}
Mandelbrot$^{\ref{Mandelbrot56}}$ was probably the first  to link
statistical physics with the theory of statistical inference.  
He obtained  a thermodynamic uncertainty relation using the methods of
estimation theory.
 
Like Rosenfeld and Landau and Lifshitz, he adopts the point of view of 
statistical thermodynamics in which the concept of an underlying 
microscopic phase space is superfluous and one works directly with probability 
distributions over macroscopic variables of the system.
  In contrast to the previous discussion, where such a distribution is
obtained from the Einstein postulate, one now starts by assuming the
existence of a probability distribution $p_\beta(U)$ to describe the
energy fluctuations of a system in contact with a heat bath at temperature
$\beta$. 
 The temperature of the system is thus represented by a parameter in its
probabilistic description.  By imposing a number of axioms
Mandelbrot$^{\ref{Mandelbrot62}} $ is able to determine the form of this
probability distribution. 
    The most important of these is the demand that sufficient estimators
for $\beta$ should be functions of the energy $U$ alone. 
    This enables him to invoke the  Pitman-Koopman theorem (\ref{PK})
which eventually leads to the form:
\be 
   p_\beta (U)  = \frac{e^{-\beta U} \omega(U)}{Z(\beta)}, \label{can}
\ee
 with $Z(\beta)$ 
the partition function, i.e.\ the normalization constant  and  $\omega(U)$ 
the so-called structure function of the system.$^{\ref{structure}}$
     
     Mandelbrot considered the question of estimating the unknown 
parameter $\beta$  of this system by measurements of the energy. 
The Fisher information in this case equals
 \be
   I_F(\beta)=(\Delta_\beta U)^2=\midd{U^2}_\beta-\midd{U}_\beta^2.
\label{IF}
\ee
 Thus, if we apply the Cram\'er-Rao inequality for  unbiased estimators
$\hat{\beta}$, we immediately find the result: 
 \be
   \Delta_\beta U \Delta_\beta \hat{\beta} \geq 1. 
   \label{CRcan} \ee 
 This is Mandelbrot's uncertainty relation between
energy and temperature. It expresses that the efficiency
with which temperature can be estimated is bounded by the spread in
energy. Note that this does not mean that the temperature fluctuates: it
is assumed throughout that the system is adequately described by the
canonical distribution function (\ref{can}) with fixed $\beta$.  Rather, the
estimators fluctuate (i.e.\ they are random quantities).  Their standard
deviation is employed, as usual in estimation theory, as a criterion to
indicate the quality (efficiency) with which $\beta$ is estimated. 
 Thus the two $\Delta$'s in the relation (\ref{CRcan}) have different interpretations,
 in contrast to the result of Rosenfeld or Schl\"ogl.

Let us make some remarks on this result.
On first sight, the use of such different interpretations 
for the two uncertainties in relation (\ref{CRcan}) may seem to reveal a 
striking disanalogy with the quantum mechanical uncertainty relations. 
However, recent work on the quantum mechanical relations has shown that here
too it is advantageous to employ concepts from the theory of 
statistical inference. Already several 
authors$^{\ref{Holevo}-\ref{Braunstein}}$ 
have advocated the 
\CR for a formulation of the quantum mechanical
uncertainty relations which improves upon Heisenberg's inequality. 
We shall discuss
these 
developments in 
section \ref{SD}. Let it suffice to remark here that 
the approach by Mandelbrot is actually in close analogy with these recent 
formulations in quantum mechanics.

Mandelbrot$^{\ref{Mandelbrot56}}$
 calls the spirit of his approach ``extremely close to that
of the conventional (Copenhagen)  approach to quantum 
theory''
and argues
that the incompatibility of quantum theory and hidden variables is
comparable to that of statistical thermodynamics and kinetic theory.
 In later works$^{\ref{Mandelbrot62},\ref{Mandelbrot64},\ref{Mandelbrot89}}$
however, he no longer claims that statistical thermodynamics 
 is incompatible with, but only indifferent to, the use of 
 kinetical or  microscopic  models.
Thus he writes: 
``Our approach\ldots   realizes a dream of the 19th century
   `energeticists': to describe matter-in-bulk without reference to atoms.
   It is a pity that all energeticists have passed away long 
ago.''$^{\ref{Mandelbrot62-1036}}$.
Indeed, his aproach can be readily extended to statistical
mechanics by assuming the existence of a  mechanical phase
space and interpreting the distribution (\ref{can}) as a marginal of a
canonical distribution over phase space:  \be p_\beta(U) =
\frac{1}{Z(\beta)} \int_{H(\x) =U}  
   e^{-\beta H(\x)} \, \dx\label{canm} \ee where 
 $(\x) \in
{I\!\!R}^{6N}$ is the microscopic state of the system and $H$ denotes its
Hamiltonian. The structure function is then identified with the area
measure of the energy hyper-surface $H(\x)=U$. 
 Of course, since the evolution of
the microscopic state is now dictated by the
 mechanical equations of motion, the validity of the
  interpretation of the probabilities as time frequencies needs additional 
  attention, e.g.\  by the assumption of an ergodic-like hypothesis.

Note, however, that such a detailed microscopic
description does not help for the estimation of $\beta$. 
 The energy is still a sufficient estimator, and therefore contains all
relevant information about the temperature. No further information is
gained by specifying other phase functions as well,
or indeed the exact microscopic state itself. 
 Thus, we can restrict our attention to the distribution over the energy.

Further, we note that, since Mandelbrot relies on the Cram\'er-Rao
inequality, his uncertainty relation is valid only when the canonical
distributions are regular (cf.\ footnote \ref{reg}).  Thus, it may fail,
for example, for systems capable of undergoing phase transitions.

Mandelbrot's result (\ref{CRcan}) applies, like those discussed in section
2, only to systems imbedded in a heat bath. Let us now ask whether it can
be extended to isolated systems in order to see whether there is some kind
of complementarity between these two contexts. The distribution function
is in this case microcanonical
\be
   p_{\epsilon}(U)=\delta(U-\epsilon),
\ee
where $\epsilon$ is the fixed  energy of the system. 

The first problem one can then raise, in analogy with the previous case,
is that of inferring the value of the energy parameter $\epsilon$.  It is
immediately clear however that
a single measurement of the energy  $U$ itself suffices to
estimate $\epsilon$ with utmost accuracy. The Fisher information is
infinite, and no informative uncertainty relation arises in this case.
That is, choosing $\hat{\epsilon} =U$ as estimator for $\epsilon$,  we
obtain: 
\be 
   \Delta_{\epsilon} \hat{\epsilon} = 0.
\ee

A next and more interesting problem is then what can be said about the
temperature in the microcanonical distribution. 
This, of course, presupposes that one can give a definition of temperature of
an isolated  system. 
Mandelbrot$^{\ref{Mandelbrot89}}$ proposes to address this problem by regarding 
the isolated system as if it had been prepared in contact with a heat bath, 
i.e.\ as if it were drawn from a canonical ensemble. In that case the 
discussion of our previous problem (i.e.\ of estimating the temperature of 
the heat bath) would have been applicable. His proposal is to treat the 
assignment of a  temperature to the  isolated system in the same way as 
the estimation of the unknown parameter $\beta$ of this hypothetical 
canonical distribution. In other words, whatever function  $\hat{\beta}(U)$ 
is a `good' estimator of $\beta$ in the canonical case is also a good 
definition of temperature in the microcanonical case. 

It is interesting, therefore, to consider some specific estimators for
$\beta$. Mandelbrot mentions three of them. The maximum likelihood
estimator $\hat{\beta}_1(U)$ is defined as the solution of the equation
\be
   \left(\frac{dp_\beta(U)}{d\beta}\right)_{\beta =\hat{\beta}_{1}}=0.
\ee
Other choices are:
\be
   \hat{\beta}_2(U)= \frac{d\ln \Omega (U)}{dU},
\ee
and 
\be
   \hat{\beta}_3(U)= \frac{d\ln \omega (U)}{dU}
\ee where
\be 
     \Omega (U)
     \equiv \int_0^U
    \omega(U') dU'.
\ee
These  functions are in fact often proposed 
as candidate definitions of the temperature of isolated 
systems.$^{\ref{already}}$
They generally yield different values for finite systems, 
but for typical systems in statistical mechanics (consisting of 
particles with finite range interactions),
they  converge for large numbers of particles.$^{\ref{Martinlof}}$
 For the ideal gas, for example, one has $\hat{\beta}_1(U)=
\hat{\beta}_2(U)=  3N/(2U)$ and $\hat{\beta}_3(U) =(3N-2)/(2U)$. 
However, in other cases, e.g. for a system consisting of  magnetic moments, 
where $\omega(U)$ is decreasing in a part of its domain,
$\hat{\beta}_2$ and $\hat{\beta}_3$
may take opposite signs, even in the thermodynamical limit. 
The problem of choosing a general `best' temperature function still 
seems to be undecided.

Even so, following  Mandelbrot's  proposal,
one can  associate a proper temperature to an isolated
system.
One might now expect, perhaps from the suggestions
of Landau and Lifshitz, or from a supposed symmetry in a complementarity
relationship, that there should be unavoidable fluctuations of such a
temperature function in an isolated system.
However, all the candidate functions $\hat{\beta}_{i}$ above
are functions of $U$, and thus remain constant in the microcanonical ensemble.
Hence one obtains 
$   \midd{\hat{\beta}_i}_{\epsilon} =
    \hat{\beta}_i(\epsilon)
$ and 
\be 
   \Delta_{\epsilon}U = 0, \;\;  \Delta_{\epsilon} \hat{\beta }_i= 0.
   \label{mcur}
\ee 
Thus, again, no uncertainty relation is obtained for 
the microcanonical ensemble. This result will clearly hold generally for 
all candidate temperature functions in Mandelbrot's proposal, since 
the postulate that $U$ should sufficient for $\beta$ implies that `good' 
(i.e.\ efficient) estimators depend on $U$ alone.
 
This is not the conclusion Mandelbrot draws, however.
 He proposes$^{\ref{Mandelbrot89}}$ to judge not only the temperature, but 
also the uncertainty in temperature from the point of view of the 
canonical ensemble from which the isolated system could have been a member.
 Also the uncertainty in energy is calculated from this counterfactual 
point of view. Thus we simply recover the canonical uncertainty relation 
(\ref{CRcan}) which is now, counterfactually, said to apply also to the 
microcanonical case. But this does not seem to be a satisfactory procedure 
to generalize the  validity of a relation. 
 
 We conclude that although Mandelbrot's approach encompasses a discussion of 
both isolated systems as well as systems in a heat bath, the result is 
still that there is no complementarity between canonical and 
microcanonical ensembles (isolation and thermal contact with a heat bath).

\subsection{The Bayesian approach: Lavenda}
Another major school of thought in statistical inference is Bayesianism. 
According to Bayesian statistics, probabilities can be attributed to 
all kinds of statements, in particular also to  other probability statements.
 Thus, one is allowed to assume that so-called prior probabilities
$\rho(\beta)$
can be attributed to the parameter $\beta$ in the canonical distribution.
Furthermore, the canonical distribution is interpreted as a conditional
probability $p_\beta(U)=p(U|\beta)$.  One is then able, by means of
Bayes's theorem, 
\BE 
   p(\beta|U )  
   &=&\frac{p(U|\beta) \rho(\beta)}{\mathcal{Z}(U)}    \label{post}  \\
   \mbox{with}~~~~{\mathcal{Z}(U)}&=&  \int
   p(U|\beta) \rho(\beta)d\beta    \label{Z}
 \EE
 to obtain a so-called posterior probability distribution over $\beta$
conditioned on a given value of $U$. The tenet of the Bayesian approach is
that all inferential judgements about $\beta$ on the basis of an observed
value of $U$ are encapsulated in this posterior distribution. In
particular, the uncertainty about its value can be quantified by the
standard deviation of the posterior distribution. 

The major problem in Bayesian statistics is the choice of the prior
distribution $\rho$. Usually this choice is made with the help of an
argument
appealing to our prior ignorance of the value of $\beta$. Therefore,
Bayesian statistics is often regarded as intimately connected to a
`degrees of belief' interpretation of probability, in contrast to the relative
frequency interpretation adopted above. 
The simplest choice for a prior distribution representing 
ignorance would, of course, be a uniform distribution. However, 
Jeffreys$^{\ref{Jeffreys}}$ has argued that in the case
of a non-negative physical quantity $\beta$, appearing as a parameter in
a probability distribution, the appropriate representation of ignorance is
by putting 
\be
   \rho(\beta) = \sqrt{I_F(\beta)} \label{Jeff}
\ee
where $I_F(\beta)$ is the Fisher information (\ref{I_F}). 
The main motivation for this choice is that this makes the 
distribution invariant under bijective reparameterization, which is 
obviously desirable for a distribution intended to represent a state of 
ignorance. A drawback is that the prior distribution (\ref{Jeff}) is often 
not normalizable. However, the posterior distribution usually is. 

 Let us now consider the question  whether 
 a thermodynamic uncertainty relation can be obtained in this approach.
That is, we ask whether there is a relation between the standard deviation
in $\Delta_U \beta$ of (\ref{post}) and the standard deviation $\Delta_\beta U$
of (\ref{can}).
 Unfortunately, in general nothing definite about this question can be said. 
Indeed, it is obvious that both standard deviations contain a 
parameter, $U$ (the observed value of the total energy) and $\beta$
respectively,
which are functionally independent. As we will see, however, it is possible
to derive uncertainty relations for specific choices of these parameters.
 
 Note that since this approach yields a probability
distribution over $\beta$, its uncertainty can be quantified by means of 
a proper standard deviation of $\beta$.
  This is possible because the Bayesian approach obliterates, 
as a matter of principle, the distinction between parameters and random
quantities. However, this does not mean that $\beta$ fluctuates; rather,
the posterior distribution from which $\Delta \beta$ is calculated has 
a meaning in terms of degrees of belief, so that
$\Delta \beta$ represents a region of epistemic uncertainty.  

Using this Bayesian approach,
Lavenda$^{\ref{Lavenda87}-\ref{Lavenda91}}$ 
claims to have arrived at an uncertainty relation: 
\be
   \Delta U \Delta \beta \geq 1,
   \label{urLavenda} 
\ee 
where the equality sign applies to equilibrium distributions, and a strict
inequality holds for irreversible processes. 
He argues that this result ``stands in defense of a purely statistical
interpretation of thermodynamics''$^{\ref{Lavenda88b-686}}$,
 just as the Heisenberg uncertainty principle protects quantum mechanics 
 from a hidden variables interpretation.
 This suggests that the uncertainty relation would forbid a mechanical
underpinning of statistical thermodynamics.
  However, he apparently does not wish to go so far, because he also
writes that the statistical inference approach to thermodynamics is
analogous to 
the orthodox approach to quantum theory in the sense that it
``circumvents a more fundamental, molecular description.'' 
This suggests that a molecular  description is merely unnecessary, 
rather than impossible; a position which  would come close to that of
Mandelbrot 
in his later writings. However, Lavenda seems to have changed his views in
the  opposite direction, since later$^{\ref{Lavenda91-6}}$
he does argue that the
thermodynamic uncertainty relation excludes a mechanical underpinning:
 ``The very fact that uncertainty relations exist in thermodynamics 
 between \ldots energy and inverse temperature, makes it all but impossible 
 that a probabilistic interpretation of thermodynamics would
ever be superseded by [a] deterministic one, rooted in the dynamics of 
large assemblies of molecules''. 

We now turn to the  derivation of (\ref{urLavenda}).
Lavenda provides several, but all based on different assumptions.
 We shall discuss two of his derivations of (\ref{urLavenda}) 
 with equality sign, and one of the corresponding inequality. 
A major role throughout the approach is played by the identification 
of the value of a maximum likelihood estimator and the expected value 
of a parameter in the posterior probability distribution. 
He is able to trace the assumption back to Gauss's attempts to justify 
the method of least squares, and therefore baptizes it ``Gauss' principle''.
However, he does not give a convincing motivation for this assumption. 
Since in principle estimators and expected values are very different
quantities, we reject this assumption, and have therefore tried to 
circumvent it as much as possible in our reconstruction of Lavenda's work. 

To obtain the uncertainty relation with equality sign, 
Lavenda$^{\ref{Lavenda88b}} $
starts
from the assumption that the system consists of 
a large number $n$ of non-interacting identical subsystems, so that its
total energy 
can be 
written as $U_{\rm tot} = \sum_{i=1}^n U_i$. Each subsystem has a
canonical distribution.  Thus, 
\be 
   p(U_1, \ldots,U_n | \beta ) = \frac{
   \prod_{i=1}^n\omega(U_{i})}{Z^n(\beta) } 
   e^{-\beta{\Ut}} \label{ll}
\ee
and our goal is to consider the posterior distribution
\be
   p(\beta | U_1 ,\ldots, U_n) = \frac{p( U_1 , \ldots ,U_n |\beta )
   \rho(\beta)  }{{\cal Z}(U)}.
\ee
We have already noted that the total energy $\Ut$ is a
sufficient estimator  for $\beta$ in the canonical distribution. 
For the Bayesian approach the cash value of this is that  the posterior 
depend on $\Ut$ alone:
\be
   p(\beta | U_1 ,\ldots, U_n) = p(\beta | \Ut ) \label{LL'}.
\ee
In order to obtain the approximate shape of this distribution 
for large $n$, one can make a second-order Taylor expansion of $\log
p(U_1,\ldots, U_n|\beta) 
$ as a function of $\beta$ around its maximum value: 
\be
p(U_1,\ldots, U_n|\beta)
 \simeq  p(\hat{\beta}_1 (\Ut))   
   \exp \left( -  \frac{1}{2} (\beta - \hat{\beta}_1(\Ut))^2 
   J(\Ut)  \right) \label{L3}
\ee
where $\hat{\beta}_1(\Ut)$ is the maximum likelihood estimate, and  
\be
   J(\Ut)=  - \left( \frac{\partial^2 \log p(\Ut|\beta) }{(\partial
\beta)^2} 
   \right)_{\beta = \hat{\beta}_1(\Ut)}.
   \label{JU}
\ee
It is easy to show that 
\be
   J(\Ut)=  - \left( \frac{\partial^2 \log Z(\beta) }{(\partial 
   \beta)^2} 
   \right)_{\beta =  n \hat{\beta}_1(\Ut)} = 
   I_{F} (\hat{\beta}_1(\Ut) ).
\ee
Thus, this is just another version of the Fisher information.

For large $n$ one expects that  $J\propto n $, so that (\ref{L3}) 
is  appreciably different from zero only in a small region around 
$\hat{\beta}_1(\Ut)$.
Inserting in (\ref{post}), and assuming that $\rho(\beta)$  behaves reasonably
smoothly and does not vanish in  this  region,  one obtains  
\be 
   p(\beta| \Ut) \simeq \sqrt{\frac{J(\Ut)} {2 \pi}}
   \exp \left( -  \frac{1}{2} (\beta - \hat{\beta}_1)^2 J(\Ut)  \right)
   \label{L4} 
\ee
so that the prior drops out of the posterior. 

The standard deviation of $\beta$ in the Gaussian posterior distribution
(\ref{L4}) is obviously 
\be 
   \Delta_U \beta= \frac{1}{\sqrt{J(U)}}.
\ee
Also, it is clear from (\ref{JU}) that 
\be
   J(\Ut)= n\left(\Delta_{\hat{\beta}_1(\Ut)} U\right)^2= 
(\Delta \Ut)^2.
\ee
Combination of these leads to the uncertainty relation 
(\ref{urLavenda}) with equality sign, 
for the specific choice $\beta=\hat{\beta}_1(U)$. 

Lavenda also offers an argument to determine the form of the prior distribution
from the asymptotic expression and the demand that the expectation 
value of $\beta$ in the posterior  distribution  (\ref{post})
should coincide with the maximum likelihood estimator for this parameter.
Interestingly, this leads to the Jeffreys prior. His argument seems to be 
erroneous,$^{\ref{erroneous}}$
but since the prior drops out anyway, this has no effect on the remainder
of his analysis. 

Lavenda also provides another derivation for the uncertainty relation with
equality sign.$^{\ref{Lavenda88a},\ref{Lavenda91}}
$
Here, no assumption about the number of subsystems 
and no Gaussian approximation are needed. Instead,
Gauss' principle is invoked in order to equate the expectation value of
$\beta$
in the posterior distribution with the 
parameter in the canonical distribution. 
   The entropy $S(\midd{U})$ is  identified
with $- \ln \mathcal{Z}(\midd{U})$, the second derivative of which equals
\be
   \pd{^2 S}{\midd{U}^2}=-\left( \Delta_{\midd{U}}\beta \right)^2.
\ee
Here $\midd{U}$ denotes the expectation value of the energy in the
canonical
distribution.
On the other hand, by equating $\midd{\beta}_{\midd{U}}$
with $\beta$,
\be
   \pd{^2 S}{\midd{U}^2}=
   \pd{\midd{\beta}_{\midd{U}}}{\midd{U}}
   =\left( \pd{\midd{U}}{\beta}\right)^{-1}
   =-\left( \Delta_\beta U \right)^{-2},
   \label{LavGau}
\ee
and the desired result follows. Again the uncertainty relation is derived 
for a specific value for one of the parameters only, 
but now for $U=\midd{U}$ instead of $\beta=\hat{\beta}_1$. 

Let us now consider relation (\ref{urLavenda}) with inequality 
sign. Lavenda notes that such an inquality  is connected with the 
Cram\'er-Rao inequality, but claims that its physical content lies in 
the existence of irreversible processes.  
His idea is to make use of the Second Law, $\Delta S_1+\Delta S_2 \geq 0$, 
for the entropy increase during a process in an isolated compound
system, and transform this inequality into the desired uncertainty relation
(i.e.\ (\ref{urLavenda}) with inequality sign) by making use of 
standard thermodynamic relations and certain identifications of thermodynamic
quantities with statistical ones.

However, as we shall show, the derivations provided by Lavenda are in error:  the validity of the uncertainty relation with equality sign is 
presupposed in the statement of the relation with inequality sign. 
Thus, his claim that the strict inequality is instantiated by
 irreversible processes is mistaken. 

In his book$^{\ref{Lavenda91-196}}$ Lavenda
considers two subsystems initially
at temperatures $T_1$ and $T_2$ and assumes they are placed in thermal contact
so that they are allowed to exchange energy in the form of heat, 
but not  any work. Then, if an infinitesimal amount of
energy
$\delta U$ is exchanged the total entropy will change by:
\be 
   \delta S_1 + \delta S_2 =  \left( \frac{1}{T_1} - \frac{1}{T_2}\right)
   \delta U  \geq 0   \label{s1s2}.
\ee
Now suppose that both systems are ideal gases, and that
as a result of the energy exchange, system 1 changes its temperature 
from $T_1$ to $T_1 + \delta T$. Then its entropy change will be 
\be   \delta S_1  = c_V  \ln \frac{T_1 + \delta T}{T_1} \approx c_V
\frac{\delta T}{T_1} = - c_V T_1 \delta \beta \ee

Hence 
\BE \delta S_1 + \delta S_2  &=& - c_V T_1 \delta \beta  - \frac {\delta U
}{T_2}
= \left( -c_V T_1 T_2 \frac{\delta \beta }{\delta U} -1  \right)
\frac{\delta U}{T_2}\EE
Now, by some  inscrutable reasoning, Lavenda argues for  the validity of the following 
relations   (at least valid up to second order)  
\be  I_F  =  - \frac{\delta U}{\delta \beta} = \Delta U^2 = 
(\Delta\beta)^{-2} = 
c_V T_1 T_2  \label{mys} \ee
where $I_F$ is the Fisher information.
Using these equations one finds 
\be\delta S_1 + \delta S_2  = \left( I_F \Delta \beta ^2  -1 \right)
\frac{\delta U}{T_2}  \geq 0 \ee
from which  the desired inequality follows 
\be 
   \Delta U  \Delta \beta \geq 1 
\ee
  since $\delta U >0$.   Thus, according to Lavenda, the uncertainty relation
actually stems from the Second Law of thermodynamics.
 
This argument is obviously erroneous.  A first objection is that the
validity of the left-hand part (\ref{s1s2}) is already confined to the
case of reversible proceses only, so that the total entropy must remain
constant.  What is worse, however, is that the relations
 (\ref{mys}) already by themselves imply the stronger relation
 $\Delta U \Delta \beta = 1$, regardless of whether we consider a
reversible or irreversible process. 

In conclusion, there is no indication that an uncertainty relation with
inequality sign can be derived in the Bayesian approach, let alone be
explained in terms of irreversible processes. Further, we have seen that
only one of the derivations Lavenda presents for the relation with
equality sign, can stand the test of critical analysis. This leads to the
result that for a system consisting of a large number of subsystems, and
with a Gaussian approximation for the posterior distribution, the relation
 \be
   \Delta_{\hat{\beta}_1(U)}U \Delta_U \beta =1
   \label{urL2}
\ee
 holds asymptotically.

One cannot help but note the close mathematical connection between this 
 derivation and that of Mandelbrot, in spite of their widely different
statistical philosophy. We have seen that in the large sample approximation,
the Bayesian prior drops out of equation (\ref{L4}) and the posterior 
becomes simply proportional to the likelihood function 
$L_U(\beta) =p_\beta(U)$. 
On account of the symmetry between $\beta$ and $\hat{\beta}$ 
in this Gaussian, it becomes formally immaterial whether we regard
this  expression as a distribution over $\beta$ (as favoured by
Lavenda) or over the estimator $\hat{\beta}_1$, as done by Mandelbrot.
The standard deviations will in both cases be given by $1/\sqrt{I_F}$.
Thus, Lavenda's result can be seen as a consequence of Mandelbrot's
inequality (and the fact that the Maximum Likelihood estimator
asymptotically saturates the Cram\'er-Rao bound). 
 
 \subsection{The  approach by fiducial probability: Lindhard}
 A different approach to thermodynamical uncertainty relations was given by
Lindhard$^{\ref{Lindhard}}$.  Like Rosenfeld, Lindhard restricts his
discussion to a system with fixed volume in contact with a heat bath. Like
Mandelbrot and Lavenda, he uses Gibbsian ensembles to determine the
probability distributions, instead of the Einstein postulate.
   But unlike previous authors, Lindhard considers both the canonical
and microcanonical ensembles as well as intermediate cases, describing a
small system in thermal contact with a heat bath of varying size. 

  The relation  he derives is
 \be
   (\Delta U)^2 + C^2(\Delta T)^2 = kT^2C.
   \label{urLindhard}
\ee
Here, $\Delta T$ and $\Delta U$ are  standard deviations of temperature
and energy of the system,
and $C=\partial U / \partial T$ is its heat capacity.
This relation has a somewhat different appearance from the uncertainty 
relations we have encountered so far, but it  is still an uncertainty
relation in the sense that it expresses that one standard deviation can only
become small at the expense of the other's increase. 
 
The relation is intended to cover as extreme cases both the canonical
ensemble, where, according to Lindhard,  
$ \Delta T =0$ and 
\be(\Delta U)^2 = kT^2 C\label{Lc},\ee
 and the microcanonical ensemble, for which
$\Delta U= 0$ and  
\be (\Delta T)^2 = kT^2/C.\label{Lmc} \ee
Thus, we here have a candidate  relation which holds for a class of ensembles,
and can be seen as expressing  a complementarity, not only between 
temperature and energy, but also between the canonical and 
microcanonical description, in the same way as the (improper) eigenstates of 
position and momentum appear as extreme cases in the quantum mechanical 
uncertainty relations.

To obtain his result Lindhard starts from the canonical distribution 
(\ref{can}) describing a system in contact with an infinitely large heat bath
at fixed temperature $T$. The standard deviation of its energy can
be expressed as 
\be 
    (\Delta U )^2 = kT^2 \pd{\midd{U}}{T} =kT^2 \midd{C}.
\ee 
This yields (\ref{Lc}).  Next, Lindhard ``inverts'' the canonical
distribution, to obtain \be
   p_U(\beta)=\pd{}{\beta}\int_0^{U}p_{\beta}(U')dU',
   \label{inv}
\ee
 and interprets this as the probability of the unknown temperature of the
heat bath.  He argues that, due to its infinite size, the heat bath can
itself be regarded as an isolated system, and can thus be described by
means of a microcanonical distribution.  Hence the inverted distribution
(\ref{inv}) can be interpreted as a microcanonical probability
distribution for temperature.  The standard deviation of this distribution
should then yield (\ref{Lmc}). Lindhard does not prove this, but notes
that it is approximately true when the heat bath is an ideal gas.

To get to intermediate distributions, Lindhard observes that, just like
a canonically distributed system is in contact with a heat bath of 
infinite heat capacity, an isolated system can be construed as being 
in contact with a heat bath of zero heat capacity.
Thus one obtains  intermediate cases by considering a heat bath
having a capacity $\xi C$, with $C$ the capacity of the system itself
and $ 0<\xi< \infty$.
Lindhard now assumes that  the probability distribution for 
the energy of such a system  takes  the shape of  a Gaussian
 distribution with width $(\Delta U)^2=\sigma_c^2\xi/(1+\xi)$,
 where $\sigma_c^2$ is the  standard deviation  in the energy for a canonical
distribution (\ref{Lc}). 
Combining this result with  (\ref{Lmc}) which now reads
\be  (\Delta T)^2 = \frac{kT^2 }{(1 + \xi) C}\ee
and eliminating $\xi$ we finally obtain the result (\ref{urLindhard}).
Lindhard also claims, without proof, that for more general probability 
distributions this result holds too, with the equality replaced by a 
greater or equal sign.

There are, obviously, a number of objections to Lindhard's approach.
On first sight, his inversion procedure seems obscure. 
It is interesting to note, however, that the formula (\ref{inv})
for a probability distribution over an unknown parameter in the light of an
observed value $U$ is well-known in another approach to  statistical inference
 proposed by Fisher and usually  called  fiducial probability.
 This approach provides a rival statistical procedure,
alternative to both estimation theory and Bayesian statistics, and gives
Lindhard's inversion technique a theoretical background.
 Fisher restricted this fiducial argument to the
cases where a sufficient estimator for the parameter exists and 
where its distribution is monotonous as a function of the parameter. 
 This approach,  however, is controversial and plagued by
paradoxes.$^{\ref{Hacking.e.a}}$
 Mandelbrot$^{\ref{Mandelbrot56-193}}$
 captures general opinion in the remark
that the fiducial argument is ``often regarded as to be carefully
avoided''.

Secondly, there are many gaps in Lindhard's argument. It is not clear that
the inverted distribution will indeed have the standard deviation
(\ref{Lmc}) for systems other than an ideal gas. Also, his description of
the intermediate cases by means of Gaussian approximations seems to be too
simplified to be persuasive.

Remarkable is finally that Lindhard's argument involves a shift in
the system under consideration. The distribution (\ref{inv}) pertains, in
first instance, to the heat bath. It is used, however, to describe 
the small system.  In order to make this shift Lindhard simply assumes that the
temperature fluctuations of the total system equal those of its subsystems.
This is in marked contrast with all other authors on the subject. In fact,
there are also experimental indications against the validity of this
assumption.$^{\ref{Chui}}$

\section{Statistical distance \label{SD}}
We now describe a fourth approach to statistical inference, which we favour. 
This is the likelihood approach, which was also developed by Fisher, and later
advocated by Barnard, Hacking and Edwards.
We shall show how this leads to a natural measure of inaccuracy in a parameter
and use this for the formulation of uncertainty relations,  both quantum
mechanical and for temperature-energy in the canonical distribution.  
  
  The idea is here, first of all, that  the likelihood function itself conveys
all information provided by the data about the unknown parameter. 
 In this respect the approach agrees with Bayesianism.
 But now the value of the logarithm of the likelihood function is interpreted 
 as the relative support that the data bestow on parameter values.
 Thus, the parameter value for which the likelihood is maximal is regarded as 
 the best supported one. This is in close agreement with 
 the use of ML estimators in estimation theory. The basic 
 difference with estimation theory is that the quality  of the inference is
  judged not by the standard deviation of the estimator but by the form
 of the likelihood function. 

Let us write 
\be 
   S_x(\theta) := \ln p_\theta(x) 
\ee
for the support function and assume for the moment that this is a smooth 
function of $\theta$.
 A natural way to quantify its width is then by the curvature at its maximum.
 Hence, 
\be 
     \left. - \frac{d^2}{d\theta^2} \log p_\theta(x)\right|_{%
     \theta=\theta_{\rm  max}}
\ee
gives a measure of the uncertainty in the values of $\theta$ on the basis of
observed data $x$.

Now take a slightly more abstract point of view and consider 
the inferences to be expected if the data $x$ are drawn from a probability 
distribution $p_{\theta_0}$.  Then, the expected support becomes
\be 
   S_{\theta_0}(\theta)= - \int p_{\theta_0}(x)  \log p_\theta(x) dx
\ee
with a maximum at  $\theta=\theta_0$.
The expected width is then
\be 
   \left.- \int p_{\theta_0}(x)  \frac{d^2}{d\theta^2} \log p_\theta(x)
    \right|_{\theta=\theta_0} dx
\ee
which is just another form of the Fisher information (\ref{I_F}).
We thus see that in the likelihood approach 
this expression serves not merely as a theoretical bound for the efficiency 
of all estimators, but has itself a definite statistical interpretation. 
It represents the width of the expected support function,
i.e.\ how easily $\theta_0$ can be distinguished from slightly 
different parameter values. In this sense it is really a measure of
how much information one may expect to obtain about
the parameter from an observation.
 
It has been noted by many authors (Rao, Jeffreys) that the Fisher information
actually defines a metric on the set of all parameter values $\theta$.
That is 
\be 
  \delta d = \frac{1}{2}  \sqrt{I(\theta)}\, \delta \theta \label{ds}
\ee
provides an distance element  for the family $\{p_\theta\}$ which is
invariant under parameter transformations.
In fact by allowing multidimensional, or even infinite-dimensional parameters,
one can extend this metric over all probability distributions on a given
outcome set. We refer to other works for
details.$^{\ref{Wootters},\ref{HU}}$
 The distance between probability distributions is given by 
 \be
    d(p_{\theta_0},p_{\theta_1})=\arccos \int \sqrt{p_{\theta_0}(x)
   p_{\theta_1}(x)} \,dx.
   \label{distance}
\ee
(In the following, we shall write $d(\theta_0,\theta_1)$ for short.)
This then provides a measure of distinguishability of two probability
distributions. It is not only invariant under reparametrization, but even 
completely independent of the original family with which we started. 
Thus it remains useful also when this family is not smooth.
This use of the Fisher information in the likelihood approach as a metric
expressing the distinguishability of distributions should not be confused 
with Jeffreys' proposal to use it as a prior probability distribution.

It is possible to give a lower bound for the right-hand side in
terms of the endpoints $p_{\theta_0}$ and $p_{\theta_1}$ only 
(see the Appendix), to wit:
\be
   d(p_{\theta_0},p_{\theta_1})\geq\arccos\left[ \left(
   1+\frac{a^2}{(2 \Delta \hat{\theta})^2}   \right)^{-1/2}\right],
   \label{CRint} \ee where
$a=(1/2)\,|\midd{\hat{\theta}}_{\theta_0}-\midd{\hat{\theta}}_{\theta_1}|$,
and $\Delta\hat{\theta} = \max(\Delta_{\theta_0} \hat{\theta},
\Delta_{\theta_1} \hat{\theta})$.  This relation illustrates the fact
that when the distance between two probability distributions is small, the
inefficiency of any estimator is necessarily large. 
We can put this in a more transparent form by defining an inaccuracy in 
$\theta$, 
as the smallest  parameter difference 
needed to produce a statistical distance greater than $\alpha$, 
where $\alpha$ is some convenient number  between $0$ and $\pi/2$.
Thus, we define  $\delta_\alpha\theta$ as the smallest positive solution
of
\be
   d(\theta, \theta+ \delta_\alpha \theta ) = \alpha \label{inac}.
\ee
Then the inequality
\be 
   {\delta_\alpha \theta }^2 
   \leq \left(
   {\cos^{-2} \alpha} -1\right)
   {\Delta \hat{\theta}}^2
   \label{impro}
\ee
shows that $\delta_\alpha \theta$ indeed provides a lower bound to
estimation
efficiency. 

Of course, if the original family is regular (i.e.\ represented by a smooth
curve), it is well approximated locally  by a geodesic, and the improvement
obtained over the \CR  is spurious. 
By taking $\delta \theta = \theta_1-\theta_0$ infinitesimal one 
easily shows from (\ref{ds}) that the inequality (\ref{impro}) reduces to
the Cram\'er-Rao inequality. 
 However, if the family is singular, the \CR
does not apply,  whereas the inequality (\ref{impro})
still yields a lower bound to the estimation efficiency.

An advantage of this approach is that exactly the same approach 
can and indeed already has been taken to the quantum mechanical
uncertainty relations.$^{\ref{HU},\ref{Uffink}}$
Indeed,  consider a set of quantum states $\ket{\psi_x}=
e^{-ixP}\ket{\psi}$ which are
 mutually shifted in space by a parameter $x$, and suppose we want to
make an inference about this parameter.
If we perform a measurement of some obervable $A$ with eigenstates
$\ket{a}$,
say, the problem becomes the comparison of probability distributions
$|\inpr{\psi_{x}}{a}|^2 $ for the unknown parameter $x$. Just as in the
classical case, we can define a statistical distance between two states:
\be d_{A}(\psi_{1}, \psi_2) =  \arccos \sum_a  |
\inpr{\psi_1}{a}\inpr{a}{\psi_2}|
\ee
The only important distinction with the classical case is that 
in quantum theory this statistical distance depends on the choice of the
 observable $A$.   It is natural then to introduce the \emph{absolute}
statistical distance as the supremum over all observables:
\be d_{\rm abs}(\psi_{1}, \psi_2) = \sup_{A}
d_A(\psi_{1}, \psi_2) =  \arccos | \inpr{\psi_1}{\psi_2}| .\ee

If we define an inaccuracy  $\delta_\alpha  x $
 in the location parameter $x$ just as before, i.e.\ 
 as the smallest value  of  $\delta x$  that solves
  \be 
d_{\rm abs}(\psi_{x}, \psi_{x+\delta x}) =  \alpha \ee
 one obtains the relations
 \be 
\delta_\alpha  x 
\Delta P \geq \hbar  \alpha
\label{xp}\ee
where $\Delta P$ is the  standard deviation in the momentum observable of
the quantum system.
Thus, the inacuracy in the location of the state  in space
is related to the standard deviation in its momentum.

Similarly, suppose that one wants to estimate the age 
of a quantum system, i.e.\  to estimate the parameter $t$ in its 
evolution
$\ket{\psi _t } = E^{iHt/\hbar} \ket{\psi}$, 
where $H$ denotes the Hamiltonian. Then one has
\be 
   \delta_\alpha  t \Delta H \geq \hbar \alpha. \label{Et}
\ee
For an unstable system, such as a decaying atom, this gives the well-known relation
between  half-life and line width.

Taking  $\delta x$ (or $\delta t)$  infinitesimal, these relations reduce
to the Cram\'er-Rao version:
\be 
   \Delta_x \hat{Q} \Delta_x P \geq \frac{1}{2} 
   \left| \frac{d\midd{\hat{Q}}}{dx}\right| \label{crv1}
\ee
\be 
   \Delta_t \hat{T} \Delta_t H \geq \frac{1}{2} \left|
   \frac{d\midd{\hat{T}}}{dt}
   \right|\label{crv2}
\ee 
for any  observable  $\hat{Q}$ (or $\hat{T})$  which would be useful 
for estimating the location (or age) of the system.  
Relation (\ref{crv1}) was given by
L\'evy-Leblond$^{\ref{LLB}}$.
When  applied  to an `unbiased' observable, i.e.\ when  
\be\
   \midd{\hat{Q}}= \bra{\psi_x }  \hat{Q} \ket{\psi_x} = x 
\ee
we obtain the  standard Heisenberg form of the uncertainty relation, 
although for a more general class of operators than  position $Q$ 
alone.
 The last-mentioned
version of the uncertainty relation  (\ref{crv2}) for energy and time was
already derived by Mandelstam and Tamm$^{\ref{MT}}$
   
Let us now apply the concept of statistical distance and 
the resulting generalized uncertainty relations to statistical mechanics. 
The distance between two canonical distributions is 
\be
   d(p_{\beta_1},p_{\beta_2})=\arccos \left[ \frac{Z(\frac{\beta_1+\beta_2}
   {2})}{\sqrt{Z(\beta_1)Z(\beta_2)}} \right].
\ee

This relation is interesting in its own right. 
It is a well-known, and oft-repeated fact that the canonical partition
function $Z(\beta)$ is log-convex. This is equivalent to the 
the statement that the above factor between brackets is always 
less than or equal to one.
The fact that its arccosine forms a distance function does not seem to be 
so well known. 
In this case we obtain \be
    \delta \beta  \overline{\Delta U} \geq \alpha
  \ee
   with
   \be
\overline{\Delta U} = \frac{1}{\delta \beta}\int_\beta^{\beta+\delta \beta}
\Delta_\beta U  d\beta\label{av}\ee 
in analogy with (\ref{xp}) and (\ref{Et}). 
  Thus the symmetry in these two types of uncertainty relations is recovered. 
The only difference with the quantum mechanical
relations is that for the canonical family
$\Delta U$ depends on $\beta$, whereas for a free quantum system 
$\Delta P$ and $\Delta H$ do not depend on $x$ or $t$, which explains why 
an average as in  (\ref{av}) is unnecessary.

 \section{Interpolation between canonical and microcanonical ensembles;
the case study by Prosper}

Up till now we have met several unsuccessful arguments aiming to extend
the uncertainty relation for energy and temperature from the canonical
ensemble to the microcanonical ensemble. This raises the question whether
there actually is a complementarity between these two ensembles, as
envisaged by Bohr or Landau and Lifschitz.  It is therefore of interest to
analyze thermal fluctuations in intermediate situations. 

 A case study of such thermal fluctuations in a classical ideal gas
 was given by Prosper$^{\ref{Prosper}}$.  Consider a system in $d$ spatial
dimensions, consisting of $N$ particles in contact with a another system
of $M$ particles acting as a finite size
 heat bath.  It is assumed that both are ideal gases. 

We assume that the total system, consisting of $N+M$ particles, is 
described by a microcanonical distribution with fixed energy $\epsilon$.
From this one can calculate the probability that the system  has an energy
$U$:
  \be
    p_{\epsilon}(U) = \frac{1}{B(n,m)} \left( \frac{U}{\epsilon}
   \right) ^{n -1} \left( 1-\frac{U}{\epsilon}\right)^{m-1}   
   \frac{1}{\epsilon},    \;\;\;\;  0\leq U \leq \epsilon  \label{nm}
 \ee
where $ n=  \frac{d}{2} N$; $m= \frac{d}{2}M$; $N,M \geq1$  and
 \be  B(n,m) = \frac{\Gamma(n) \Gamma(m)}{\Gamma ( n+m)}. \ee
In the limit $m\rightarrow \infty,\frac{m}{\epsilon} = \beta_\infty$ 
 this distribution tends to the canonical one,  with the temperature parameter 
$\beta_\infty $. 
One the other hand, when the size of the heat bath is negligible compared to
the system, i.e.\ when  $m$  becomes very small  
compared to $n$, the distribution becomes sharply peaked just below the value
$U=\epsilon$,  resembling the microcanonical one.
(Note, however, that the representation (\ref{nm}) ceases to be valid 
for  $m =0$.)
Thus, for varying $m$ we have a family of distributions that interpolate
between the canonical and microcanonical ensembles.

The standard deviation of the energy is 
\be
   (\Delta U)^2=\frac{nm\epsilon^2}{(n+m+1)(n+m)^2}
,  \label{Ugas}
\ee
and the question is again what to say about the temperature of the system.
Prosper uses a Bayesian approach to quantify its uncertainty.
However, as we have seen in the work of Lavenda, this uncertainty in
temperature  will not lead to an uncertainty relation for a finite system.
 Instead, we shall try Mandelbrot's approach and compare this to the
statistical distance approach.   

We first assign a temperature-like parameter to the system by reparameterizing
$\epsilon$. The choice of such a parameter is only straightforward 
in the limiting case $m\rightarrow \infty$. But for finite $m$
 the choice is more or less arbitrary. Let us put
$\beta :=  {n \over \midd{U} } =  {n+m \over \epsilon}$.
Thus this  parameter coincides with  the canonical temperature
$\beta_\infty$ as $m \to \infty$. 

Suppose we wish to estimate $\beta$ from a measurement of $U$.
The Fisher information in the parameter $\beta$ is:
\be 
  I_F(\beta)  =  \left\{\begin{array}{lll} 
  \displaystyle  
     \infty &   m =1/2,3/2,2  \\
     \frac{n^2}{\beta^2} & m=1\\
     \frac{n(m+n-1)}{(m-2 )\beta^2}  & m  >2
  \end{array} \right.
\ee
 Using (\ref{Ugas})
we obtain the \CR for the parameter $\beta$:
 \be
   (\Delta U)^2(\Delta \hat{\beta})^2 \geq 
   \frac{m(m-2)}{(n+m)^2-1}~~~~~~~(m > 2). \label{63}
 \ee
 showing the limited efficiency of all unbiased estimators for $\beta$. 
This lower bound is independent of $\beta$ and increasing in $m$. For
$m\rightarrow \infty$, it reduces to the canonical value 1, already
obtained in (\ref{CRcan}). 
 Thus for the ideal gas we see indeed a gradual transition from the
canonical case (relation (\ref{CRcan})) for $m\rightarrow \infty$ to the
microcanonical case (relation (\ref{mcur})), where the uncertainty product
vanishes. 
 Somewhat disappointing is, however, that we cannot carry out a 
limit $m \rightarrow 0$. 
Already for $m=2$, the Fisher information is infinite.
This is, however, no indication that the parameter 
$\beta$  becomes  perfectly estimable for small $m$.
Rather,
for $m\leq 2$ the distributions are singular and  the \CR no longer holds. 
(Cf.\ footnote~\ref{reg}.)

Thus the inequality (\ref{63}) expresses the gradual transition between 
canonical and microcanonical distribution, but not perfectly.
 
Here, the method of statistical distance offers a more detailed analysis
of the situation. The statistical distance  for some values of $n$ and $m$
are as follows: 
\BE 
d(\beta_0, \beta_1) & =&   \arccos I_{mn}(b)~~~~~~\mbox{with }\;\;\;\;  
    b    = \sqrt{\frac{\beta_0}{\beta_1}}~~~~~(\beta_0 \leq \beta_1)
\\
\mbox{and } \;\;\;\;
I_{11}(b)& =&b    \\
I_{21} (b)& = &\frac{1+b^2}{2b  }  + \frac{(b^2 -1)^2}{4b^2}
 \log \frac{1-b}{1+b}  \\
I_{31}(b) &=&  \frac{b}{2} (3- b^2) \\
I_{41}(b) &= &\frac{1}{32b^4}\left( 
-6 b + 22 b^3 +22b^5- 6b^7 - 3 (b^2 -1)^4   \log \frac{1-b}{1+b} \right)\\
I_{51}(b) &=& \frac{b}{6} (10 - 5b^2  + b^4).
\\
I_{1n} & = & b^n 
 \EE

\begin{figure}
  \begin{center}
\begin{picture}(288,255)(10,20)
\put(294, 0){\large $ \displaystyle\frac{\delta\beta}{\beta_0}$}
\put(-6,252){\large $ d(\beta_0 ,\beta_0+ \delta \beta)$}
  \epsfig{file=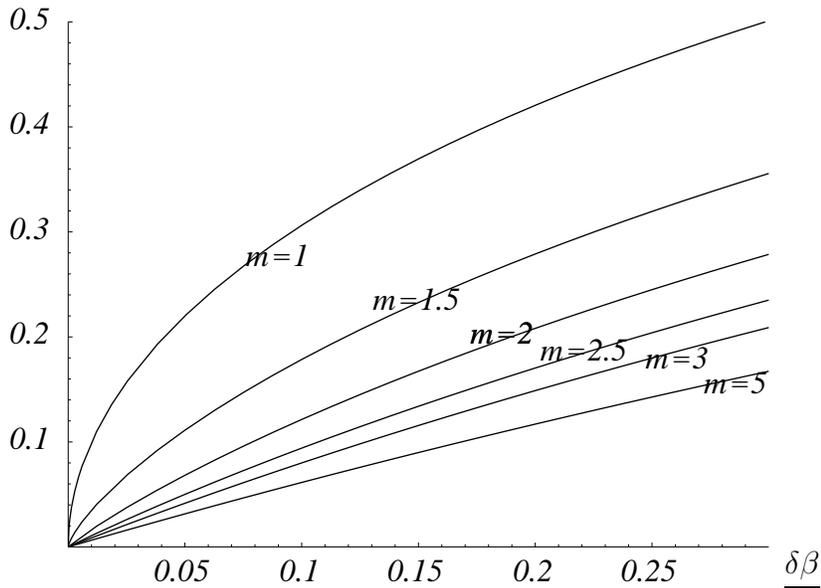}
\end{picture}
  \end{center}
  \caption{
  Statistical distance as a function of the
relative temperature difference
  $ (\delta\beta)/\beta_0$
 for $n=1$ and various values of $m$. For $m>2$ this distance 
increases linearly for small values
of  $\delta \beta$; for $m\leq 2$  the distance element is singular and 
  grows more rapidly. }
 
\end{figure}

 Figure 1 shows for the simplest case of $n=1$ how these distances behave
as a function of $\delta \beta = \beta_1 -\beta_0$.  It is seen that
for small $\delta \beta$
 the statistical  distance grows linearly
with $\delta \beta$ 
 when $m>2$, but much faster
when $m \leq 2$, due to the singularity of the associated distributions 
(\ref{nm}).
 Thus, with the same choice for the value of $\alpha$, the
inacurracy $\delta_\alpha \beta$ is much smaller for $m\leq2$ than for $
m>2$. However this does not mean that there is no lower bound. 
 For example, for $n=m=1$ one finds
  \be
    \delta_\alpha \beta \Delta_\beta U  \geq (\cos^{-2} \alpha
-1)/\sqrt{3}.
\ee

For larger systems ($n>1$), the expression for the statistical distance
becomes generally
 more
complicated. However, the same trend is observed:  if the heat
bath is small ($ m \leq 2$), the
statistical distance
  $d(\beta, \beta +\delta \beta )$ increases more rapidly than $\delta
\beta$
 but  it behaves regularly 
(i.e.\ $d(\beta, \beta +\delta\beta )\propto \delta \beta$) for $m>2$.
However  there is a positive  lower bound  to the uncertainty 
product in all cases, viz.\
\be
    \delta_\alpha \beta \Delta_{\beta} U  \geq  \sqrt{\frac{nm}{n+m+1}}
    \left(I_{mn}^{\rm inv}(\cos\alpha)^{-2} -1\right).
\ee
If we choose  $\alpha \ll 1$,  the right-hand side is here of  order
$ \alpha$ for $m>2$ and 
${\alpha}^2$ for $m=1$. 
We recover, therefore, the same conclusion as before: 
in a gradual transition from a canonical to a microcanonical distribution 
the uncertainty product is bounded by a constant gradually approaching
zero.
We thus see that the uncertainty relations obtained in this 
approach remain valid  also for singular families. However, the
right-hand side is then  much lower than in the regular case. 

Of course these conclusions are obtained only for the case study
of the ideal gas. 
Yet it remains remarkable that, in the approach using the statistical
distance, one obtains non-trivial uncertainty relations for regular as
well as singular families. A serious challenge would then be to apply this
to systems capable of phase transitions. This, however, falls outside the
scope of this paper. 

\section{Discussion}
We have reviewed several approaches to the formulation of thermodynamic
uncertainty relations in the existing literature. Only two have a
reasonably general derivation, free from undesirable simplifying
assumptions. The formulation of Schl\"ogl pertains to a version of
statistical thermodynamics founded on the Einstein postulate.  The second
formulation is by Mandelbrot. His result is valid for canonical ensembles
both in his own axiomatic version of statistical thermodynamics as well as
in statistical mechanics. In Schl\"ogl's treatment, the system has a
randomly fluctuating temperature, in Mandelbrot's case the temperature is
identified with a parameter in its probability distribution. Its
uncertainty is identified with estimation efficiency and his uncertainty
relation is valid for regular families, i.e.\ barring phase transitions. 
We have also proposed an extension of Mandelbrot's approach by means of
the theory of likelihood inference. Here the uncertainty in the
temperature parameter is  quantified by means of statistical distance.  The
uncertainty relation obtained in this way is valid also for non-regular
families. 

Both Mandelbrot's result and the generalization we proposed are in close 
analogy with formulations of uncertainty relations in quantum mechanics. 
This provides a  good reason to take them just as serious as the quantum 
mechanical relations. 
 Still, there are deep differences between statistical mechanics and
quantum theory in particular in their interpretations of probability. When
in quantum mechanics a system is described probabilistically, this is
usually said to represent the state of the system completely; in contrast,
the probability distributions in statistical mechanics are regarded as
`only' convenient tools.  In this case the mechanical phase space provides
the underlying variables which in quantum mechanics are regarded as hidden
or even non-existent.
 The probabilistic description gives only a small part of the information 
 concerning the system that could in principle be obtained. 
 But --- it also gives something more.  As we have seen, complete knowledge 
of the microstate would not suffice to infer the temperature of the system. 
What we have added by giving a probabilistic description is the notion of 
the ensemble, of which the system of interest is only a member. 
That is, we have added something which goes beyond the particular system we 
are studying.  This points to a common feature between the quantum case and 
the statistical mechanical case. Once we have accepted that our system is 
described by some probability distribution out of a certain parameterized
class, the problem of statistical inference of the parameter occurs in exactly 
the same way, and the uncertainty in this parameter can be quantified with the 
same methods.
 
In all the approaches mentioned, there is no uncertainty relation for
energy and temperature for an energetically isolated system. Therefore,
the results do not justify a claim for complementarity between energetic
isolation and thermal contact, as envisaged by Bohr and Heisenberg. 
Indeed, we conclude that no such complementarity exists. 
Mandelbrot's attempt to give his relation a more general validity by
regarding the isolated system, counterfactually, as a case which could
have been described by a canonical ensemble is not convincing.
We also come to the conclusion that there doesn't exist a 
complementarity between the microcanonical and canonical ensembles. In
particular, it is not true (as Lindhard claims) that the extremes of any
valid uncertainty relation are covered by these ensembles respectively.
In fact, the study of intermediate cases reveals that the product of the
uncertainties in energy and temperature gradually changes from zero
(microcanonical case) to one (canonical case). 

In this respect there are strong disanalogies  between the thermodynamic and 
the quantum mechanical uncertainty relations.
It is sometimes argued$^{\ref{Mandelbrot56},\ref{Lavenda88b}}$ 
that the mathematical 
basis of the uncertainty relationship in quantum mechanics (the Fourier 
transform  between position and momentum eigenstates)
is so analogous to the Laplace transform relationship 
between the canonical and microcanonical ensembles, that an analogous 
relationship should be expected. However, this analogy is misleading.
 The canonical distribution is a convex mixture of microcanonical
ones, but not vice versa. The grand-canonical distribution, where the number
of particles is also variable, is in turn a
mixture of canonical ones, etc. Thus, these ensembles of statistical mechanics 
are ordered  in a hierarchical scale of increasing randomness. 
This is quite different from the symmetry in
the Fourier relationship between the position and momentum eigenstates of
quantum mechanics.

We finally return to what seems the philosophically most surprising aspect
of our review. Do thermodynamical uncertainty relations entail an obstacle to a
microscopic underpinning of thermodynamics, in the same way as their quantum
mechanical counterparts forbid the existence of hidden variables?
We have seen that all of the protagonists in our discussion (except Lindhard)
claimed this to be true. If so, it would provide an example in classical 
physics of a situation which  is usually seen as exclusive to the quantum 
world and undreamt of in classical physics.

The first point to make in this connection is that already in quantum 
mechanics the uncertainty relations (in the standard form) do not forbid 
hidden-variables reconstructions. A much more formidable obstacle is the 
theorem of Kochen and Specker.
Hence, one cannot argue from a supposed analogy here, since the analogy 
fails already in quantum mechanics. 
The Copenhagen viewpoint that the uncertainty relations  
prevent a hidden variables reconstruction of quantum mechanics 
is due to the additional assumption 
that the physical description by quantum mechanics is already complete. 

This is not to say however, that a mechanical underpinning
or `reduction' of thermodynamics to statistical mechanics is straightforward.
  Indeed, the uncertainty relations studied here may shed new light on the
often heard statement that the relation between thermodynamics and
statistical mechanics is the archetypal example of a successful theory
reduction.  Only some thermodynamic functions can be immediately
identified as functions on phase space.  It is only for these quantities
that a complete description of the microstate of the system would be
sufficient to determine them completely.  But for 
quantities
defined statistically, i.e.\  as parameters or functionals on a
probability  distribution,
 like temperature, entropy and chemical potential,  the complete
specification of the microstate will never suffice. That is what forms the
basis
for the uncertainty relations we have studied: no phase function can
exactly mimick  such  parameters (for a range of their values).
 In this particular sense, these relations do express the impossibility of
a `hidden-variables' style extension of statistical thermodynamics.
Observe that no non-commutativity is
needed for
this conclusion.

Of course  this view depends on the definition of temperature (and chemical
potential etc.) as a parameter in a probability distribution. 
Another option that we have encountered is to consider one of
the estimator functions as a definition of temperature, rather than as a
smart guess (and therefore of course drop the qualification `estimator').
Then temperature becomes a
function on phase space, irrespective of which probability
distribution we choose to describe our system.
  No uncertainty relation for such a 
temperature function has been established except for the canonical ensembles. 
But in this case we are still left with the question which function to
choose, because many can be defined which differ radically for finite
systems. 
The reduction of thermodynamics to statistical mechanics therefore still seems 
to be an open problem. 

\section{Appendix}
Here we prove  the following inequality:
for any one-parameter family of probability distributions $p_\theta$
and any unbiased estimator $\hat{\theta}$ for $\theta $ we have
\be 
   \int \sqrt{p_{\theta_1 }(x) p_{\theta_0} (x)} dx  \leq 
   \left( 1 + \left(\frac{\theta_1  - \theta_0 }{2\Delta
   }\right)^2\right)^{-1/2},  
\ee
where 
$\Delta :=  \max ( \Delta_{\theta_1}\hat{\theta}, \Delta_{\theta_2}\hat{\theta})$.
For $n$ repeated independent observations, this result obviously 
generalizes to 
\be
   \left(\int \sqrt{p_{\theta_1 }(x) p_{\theta_0} (x)} dx\right)^{n}  \leq 
   \left(1 + \left(\frac{\theta_1  - \theta_0 }{2\Delta_n
   }\right)^2\right)^{-1/2}
\ee
with $\Delta_n :=  \max ( \Delta_{\theta_1}\hat{\theta}_n,
\Delta_{\theta_2}\hat{\theta}_n)$.

The idea behind the proof is to use the Cram\'er-Rao inequality,
not for the curve formed by the 
family $p_\theta$ but for a geodesic between the points $p_{\theta_1}$ and  
 $p_{\theta_2}$.
The geodesic between $\theta_1 $ and $\theta_0$ is the
family of distributions of the form 
\be p_\alpha   =  \left(  \alpha \sqrt{p_0} +  \beta
\sqrt{p_1}\right)^2
\label{geo}\ee
where $\alpha$ varies between 0 and 1,  and $\beta$ obeys
\be 
    \alpha^2 + \beta^2 + 2\alpha \beta c =1 
\ee
\be 
  c = \int \sqrt{p_0p_1}  dx.
\ee
Although $\hat{\theta}$ need not be a good estimator for this geodesic  
family, the CR inequality is nevertheless still valid:
\be
   \sqrt{I_F(\alpha)}
   \geq  \frac{\left|
   \frac{d\midd{\hat{\theta}}_\alpha}{d\alpha}
   \right|}{
   {\Delta_\alpha \hat{\theta}}}
\ee
and we can integrate this along the geodesic. This yields 
\be   
  \label{66}
   d(\theta_0, \theta_1) = 
   \arccos \int \sqrt{p_{\theta_1 }(x) p_{\theta_0} (x)} dx  
   = \int  {1\over 2}  \sqrt{I_F(\alpha)} d\alpha
   \geq 
   \int_0^1  
   \frac{|\frac{d \midd{\hat{\theta}} }{d\alpha}|}{2\Delta_\alpha
   \hat{\theta}} d\alpha.
\ee
Since our purpose is to find a lower bound for the right hand side, we may
assume, without loss of generality,  that  $\midd{\hat{\theta}}_\alpha$ 
is monotonous in $\alpha$. Then there exists an invertible function 
  $ y(\alpha) =  \midd{\hat{\theta}}_\alpha$. Further, since the integrals
are invariant under a linear transformation $\hat{\theta} \rightarrow
c\hat{\theta} +d$ we can arrange that
$\midd{\hat{\theta}}_0 = -a$, 
$\midd{\hat{\theta}}_1 = a$.     Thus:
 \be 
 \int_0^1  
\frac{|\frac{d \midd{\hat{\theta}} }{d \alpha}|}{2 \Delta_\alpha \hat{\theta}} 
d\alpha
\geq
\left|\int_0^1  \frac{d \midd{\hat{\theta}} }
{2 \Delta \hat{\theta}} \right|  
= 
\left|\int_{-a }^{a}  \frac{d y }
{2  \sqrt{\midd{\hat{\theta}^2}_{\alpha(y)}  -y^2}} \right|.\label{int}  
\ee

The strategy is now to find an upper bound for 
$\midd{\hat{\theta}^2}_\alpha$ in terms of  $\midd{\hat{\theta}^2}_0$, 
 $\midd{\hat{\theta}^2}_1$ and $a$.
Suppose for the moment that such an upper bound  exists, so that, say,
\be \midd{\hat{\theta}^2}_\alpha \leq A^2.  \label{A} \ee
Then we easily  obtain 
\be 
   \int_{-a}^{a}  \frac{d y }
   {2  \sqrt{\midd{\hat{\theta}}_\alpha  -y^2}}
   \geq
   \int_{-a}^{a}  \frac{d y }
   {2  \sqrt{A^2 -y^2} }
   = \arcsin \frac{a}{A}. 
\ee
Combining this with (\ref{66}) we find:
\be 
   d(p_{\theta_1}, p_{\theta_2})    \geq \arcsin \frac{a}{A} 
\ee
or 
\be
   \int\sqrt{p_{\theta_0} (x) p_{\theta_1}(x)} dx \leq \sqrt{1 - 
   \left(\frac{a}{A}\right)^2}.
   \label{res}
\ee 
It remains to show that an upper bound (\ref{A}) indeed exists.

Now the expectation $\midd{\hat{\theta}^2}_\alpha$ 
along the geodesic can be written as a convex sum: 
\be
   \midd{\hat{\theta}^2 } = \alpha^2 \midd{\hat{\theta}^2}_0 + 
   \beta^2 \midd{\hat{\theta}^2}_1 +
    2 \alpha \beta c \midd{\hat{\theta}^2}_2  \label{vex}
\ee
where $\midd{.}_2$ denotes averaging with respect to the auxiliary 
probability density
\be  
p_2(x)  = {1\over c}\sqrt{p_0(x)p_1(x)} 
\ee
Its maximum value depends on which of the three expectations  
$\midd{\hat{\theta}^2}_i$ is the largest. We distinguish two cases:
(i)  $\midd{\hat{\theta}^2}_0$ or $\midd{\hat{\theta}^2}_1$ is the largest; or 
(ii)  $ \midd{\hat{\theta}^2}_2$  is the largest of the three.

In case (i) the convex sum reaches its maximum when  $\alpha$ is 1 or 0.
Clearly the argument will be similar in both cases, so let us only consider
$\alpha =1$. 
 This gives: 
  \be \midd{\hat{\theta}^2}_\alpha  \leq  \midd{\hat{\theta}^2}_1  = \Delta^2
+a^2   \ee
This is the  value of $A^2$ in case (1).  Inserting in (\ref{res}) leads to
the result of theorem.

In case (ii) the convex sum (\ref{vex}) is maximal  
 when $\alpha^2=\beta^2 = \frac{1}{2(1+c)}$. We  thus find the upper bound
 \be  \label{bijna}
   \midd{\hat{\theta}^2}_\alpha  \leq  \frac{
\midd{\hat{\theta}^2}_0 +  \midd{\hat{\theta}^2}_1 }{1+c} + \frac{c}{1+c}
  \midd{\hat{\theta}^2}_2 
\ee
and we have to find an upper bound for  $\midd{\hat{\theta}^2}_2$. 
Using  the Cauchy-Schwartz inequality and the general inequality
$ (A+ x )(B+x) \leq ( (A+B)/2 +x)^2 $ in turn  gives:
\BE
   \midd{(\hat{\theta} -r)(\hat{\theta} +r)}_2    
   &\leq&  {1\over c} \sqrt{\midd{(\hat{\theta}+r)^2 }_0 
   \midd{(\hat{\theta}-r)^2}}_1 \nn   \\
   &=& {1\over c} 
   \sqrt{(\Delta_0^2 + (a-r)^2 )( \Delta_1^2 + (a -r)^2 )}  \nn \\
   &\leq & \frac{1}{c}( D^2 + (a-r)^2)  
   \label{ie}
\EE
where $\Delta_i := \Delta_i \hat{\theta}$, and
\be
   D^2  :=   (\Delta_0^2 + \Delta_1^2 )/2.
   \label{D} 
\ee
This inequality (\ref{ie}) is valid for all $r$;  one may choose $r$ so as
to optimize
the  strength of the  bound. This is the case when $ r = a/(1+c)$
and we obtain:
\be  
   \midd{\hat{\theta}^2 }_2  \leq {D^2\over c}    + \frac{a^2}{1+c} 
   \label{des} 
\ee
which is our  upper bound for  $\midd{\hat{\theta}^2}_2$.   
Combining this with (\ref{bijna}) gives  the desired 
upper bound for case (ii): 
\be
   \midd{\hat{\theta}^2}_\alpha  \leq  D^2 \frac{2}{1+c} + a^2
   \frac{1+2c}{1+c}.
\ee
Inserting this in (\ref{res}) yields, after a bit of algebra,
\be 
   \int \sqrt{p_{\theta_0}(x) p_{\theta_1}(x)}  dx \leq  \left( 
   1 +  \frac{a^2}{D^2} \right)^{-1/2} 
\ee
which, in view of (\ref{D}), is even slightly stronger than the
announced theorem.

\pagebreak
\section*{References}
\begin{enumerate}

\item \label{Bohr}
 See: N. Bohr, \emph{Collected Works}, edited by J. Kalckar
(North-Holland, Amsterdam,
1985), Vol. 6, pp. 316--330, 376--377. 

\item \label{Pais}
 A. Pais, \emph{Niels Bohr's times, in physics, philosophy, and polity},
(Clarendon Press, Oxford, 1991).

\item \label{Mandelbrot56}
B. Mandelbrot, ``An outline of a purely phenomenological theory of
statistical thermodynamics: 1. canonical ensembles,'' IRE Trans.\
Information Theory~{\bf IT-2}, 190--203 (1956).

\item\label{Mandelbrot62}
 B. Mandelbrot, ``The role of sufficiency and of estimation in
thermodynamics,'' Ann.\ Math.\ Stat.~{\bf 33}, 1021--1038 (1962).

\item\label{Mandelbrot64}
B. Mandelbrot,
``On the derivation of statistical thermodynamics from
purely phenomenological principles,'' J. Math.\ Phys.~{\bf 5}, 164--171
(1964).

\item \label{Rosenfeld}
L. Rosenfeld, ``Questions of irreversibility and ergodicity,''
in \emph{Ergodic Theories},
 Proceedings of the
International School of Physics ``Enrico Fermi,'' Course XIV, Varenna
1960,  edited by P.
Caldirola (Academic Press, New York, 1961). 

\item\label{Lavenda87}
B. Lavenda, ``Thermodynamic uncertainty relations and
irreversibility,'' Int.\ J.\ Theor.\ Phys.~{\bf 26}, 1069--1084 (1987). 

\item\label{Lavenda88a}
B. Lavenda, ``The Bayesian approach to thermostatistics,'' Int. J.
Theor. Phys.~{\bf 4}, 451--472 (1988).

\item\label{Lavenda88b}
B. Lavenda, ``On the phenomenological basis of statistical
thermodynamics,'' J. Phys.\ Chem.\ Solids~{\bf 49}(6), 685--693 (1988).

\item \label{Lavenda91}
B. Lavenda, \emph{Statistical Physics: a Probabilistic Approach} (J. Wiley
and Sons, New York, 1991).

\item \label{Lindhard}
J. Lindhard, `` `Complementarity' between energy and temperature,'' in
\emph{The Lesson of Quantum Theory}, edited by
J. de Boer, E. Dal and O. Ulfbeck  
(North-Holland, Amsterdam, 1986). 

\item  \label{Schloegl}
F. Schl\"ogl, ``Thermodynamic uncertainty relation,'' J. Phys.\ Chem.\
Solids~{\bf 49}(6), 679--683 (1988). 

\item
H. Feshbach, ``Small systems: when does thermodynamics apply?,''
Physics Today~{\bf 40}(11), 9--11 (1987).

Ch.\ Kittel, ``Temperature fluctuation: an oxymoron,'' Physics
Today~{\bf 41}(5), 93 (1988).

 \label{Mandelbrot89}
B. Mandelbrot, ``Temperature fluctuation: a
well-defined and
unavoidable notion,'' Physics Today~{\bf 42}(1), 71--73 (1989). 

\item  \label{LL}
L.D. Landau and E.M. Lifshitz, \emph{Statistical Physics} (Pergamon Press,
London, 1959). 

\item \label{LL62}
Ref.~\ref{LL}, p.~62. 

\item \label{LL355} Ref.~\ref{LL}, p.\ 355. 

\item \label{reg}
To be precise, inequality (\ref{CR}) is valid when
   $\frac{d}{d \theta} \int \hat{\theta}(x)p_\theta(x)dx
   = \int \hat{\theta}(x)\frac{d}{d \theta}p_\theta(x)dx $,
and
   $\int \frac{d}{d \theta}p_\theta(x)dx =0$. 

\item \label{structure}
In statistical mechanics, this structure function would be
interpreted as a measure of the number of microscopic states compatible
with energy $U$.  In Mandelbrot's approach, where there is no assumption
about microscopic states, the structure function is simply left
uninterpreted. 

\item \label{Holevo}
A.S. Holevo, \emph{Probabilistic and Statistical Aspects of Quantum
Theory}
(North-Holland, Amsterdam, 1982).

\item \label{Caianiello}
E.R. Caianiello, ``Entropy, information and quantum geometry''
in \emph{Frontiers of Non-Equili\-brium  Statistical Physics}, edited by
G.T. Moore  and M.O. Scully (Plenum Press, New York 1986), p.~453-464.

\item \label{Braunstein}
S.L. Braunstein, ``Fundamental limits to precision measurements''
 in \emph{Symposium in the Foundations  of Modern Physics 1993},  edited by P.
Busch, P. Lahti and P. Mittelstaedt  (World Scientific, Singapore, 1993).

\item \label{Mandelbrot62-1036}
 Ref.~\ref{Mandelbrot62}, p.\ 1036.

\item  \label{already}
The function $\hat{\beta}_1$ can already be found in writings of
Boltzmann, and is also used by 
A.I. Khinchin, \emph{Mathematical Foundations of Statistical Mechanics}
(Dover,
New York, 1949). 
\forget{Khinchin}  
The functions $\hat{\beta}_2$ and $\hat{\beta}_3$ were introduced by
Gibbs.

\item \label{Martinlof}
A. Martin-L\"of, \emph{Statistical Mechanics and the Foundations of
Thermodynamics} (Springer, Berlin, 1979).

\item \label{Jeffreys}
H. Jeffreys, \emph{Theory of Probability} (Clarendon Press, Oxford, 1966),
3rd ed.

\item  \label{Lavenda88b-686} Ref.~\ref{Lavenda88b}, p.\ 686.

\item \label{Lavenda91-6}
 Ref.~\ref{Lavenda91}, p.\ 6.

\item \label{erroneous}
Lavenda derives $\rho(\hat{\beta}_1) \propto
      \sqrt{I_F(\hat{\beta}_1)}$, and concludes from this
       that $\rho(\beta)$ and $\sqrt{I_F(\beta)}$ are
       proportional for all $\beta$. This overlooks  that  this
       `proportionality' was obtained for one specific value
        $\beta=\hat{\beta}_1$ only.

\item  \label{Lavenda91-196}
Ref.~\ref{Lavenda91}, pp.\ 196--197.

\item\label{Hacking.e.a}
See:
I. Hacking, \emph{Logic of Statistical Inference} (Cambridge University
Press, 1965);
T. Seidenfeld, \emph{Philosophical Problems of Statistical Inference}
(Reidel, Dordrecht, 1979) 
and A.W.F. Edwards, \emph{Likelihood} (John Hopkins University Press, 
Baltimore,
1992). 

\item  \label{Mandelbrot56-193}
Ref.~\ref{Mandelbrot56}. p.~193.

\item \label{Chui}
See T.C.P. Chui, D.R. Swanson, M.J. Adriaans, J.A. Nissen and J.A.  Lipa,
``Temperature fluctuations in the canonical ensemble,''
 Phys. Rev. Lett.~{\bf 69}(21), 3005--3008  (1992). 

\item \label{Wootters}
W.K. Wootters, ``Statistical distance and Hilbert space,'' 
Phys.\ Rev. D~{\bf 23}, 357--362 (1981).

\item 
\label{HU}
J. Hilgevoord and J. Uffink, ``Uncertainty in prediction and in
inference,'' Found. Phys.~{\bf 21}, 323--341 (1991). 

\item  \label{Uffink}
J. Uffink, ``The Rate of Evolution of a Quantum State,''
Am. J.
Phys.~{\bf 61}, 935 (1993). 

\item\label{LLB}
 J.-M. L\'evy-Leblond,
 {Phys.\ Lett.\ A.}~{\bf 111}, 353-355 (1985). 

\item \label{MT}
L. Mandelstam and I. Tamm, ``The uncertainty relation between energy and
time in non-relativistic quantum mechanics,'' J.\ Physics (USSR)~{\bf9},
249--254 (1945).

\item \label{Prosper}
H.B. Prosper, ``Temperature fluctuations in a heat bath,''
Am.\ J.\ Phys.~{\bf 61}, 54--58 (1993).

\end{enumerate}

\end{document}